\documentclass[aps,prd,
 amsmath, twocolumn, amssymb,superscriptaddress,
 aps, physrev,
]{revtex4-2}

\usepackage{graphicx}
\usepackage{dcolumn}
\usepackage{bm}
\usepackage{tikz}
\usepackage{quantikz}
\usepackage{physics}
\usepackage{hyperref}
\usepackage[capitalize]{cleveref}

\crefname{equation}{Eq.}{Eqs.}
\Crefname{equation}{Eq.}{Eqs.}

\crefname{figure}{Fig.}{Figs.}
\Crefname{figure}{Fig.}{Figs.}

\usepackage{placeins}
\usepackage{mathtools}

\usepackage{csquotes}
\usepackage[normalem]{ulem}
\DeclarePairedDelimiter{\ceil}{\lceil}{\rceil}

\newcommand{\colorgate}[4]{%
  \gate[wires=#1, style={fill=#2, draw=black, #4}]{#3}%
}

\usepackage{orcidlink} 

\begin{document}

\def\myvdots{\ \vdots\ }

\title{\textbf{(2+1)D Quantum Electrodynamics at Finite Density on a Quantum Computer}}

\author{Emil Otis Rosanowski \orcidlink{0009-0002-4072-2801}}
\email{rosanowski@hiskp.uni-bonn.de}
\affiliation{
 Transdisciplinary Research Area “Building Blocks of Matter and Fundamental Interactions” (TRA Matter) and Helmholtz Institute for Radiation and Nuclear Physics (HISKP), University of Bonn, Nussallee 14-16, 53115 Bonn, Germany
}

\author{Arianna Crippa \orcidlink{0000-0003-2376-5682}}
\affiliation{Deutsches Elektronen-Synchrotron DESY, Platanenallee 6, 15738 Zeuthen, Germany}

\author{Lena Funcke \orcidlink{0000-0001-5022-9506}}

\affiliation{
 Transdisciplinary Research Area “Building Blocks of Matter and Fundamental Interactions” (TRA Matter) and Helmholtz Institute for Radiation and Nuclear Physics (HISKP), University of Bonn, Nussallee 14-16, 53115 Bonn, Germany
}

\author{Paulo Vitor Itaboraí \orcidlink{0000-0002-4956-2958}}
\affiliation{Computation-Based Science and Technology Research Center, The Cyprus Institute, 20 Kavafi
Street, 2121 Nicosia, Cyprus}
\affiliation{Deutsches Elektronen-Synchrotron DESY, Platanenallee 6, 15738 Zeuthen, Germany}

\author{Karl Jansen \orcidlink{0000-0002-1574-7591}}
\affiliation{Computation-Based Science and Technology Research Center, The Cyprus Institute, 20 Kavafi
Street, 2121 Nicosia, Cyprus}
\affiliation{Deutsches Elektronen-Synchrotron DESY, Platanenallee 6, 15738 Zeuthen, Germany}

\author{Simran Singh \orcidlink{0000-0002-9333-3925}}

\affiliation{
 Transdisciplinary Research Area “Building Blocks of Matter and Fundamental Interactions” (TRA Matter) and Helmholtz Institute for Radiation and Nuclear Physics (HISKP), University of Bonn, Nussallee 14-16, 53115 Bonn, Germany
}

\date{\today}

\begin{abstract}
In this paper, we explore (2+1)D quantum electrodynamics (QED) at finite density on a quantum computer, including two fermion flavors. Our method employs an established
gauge-invariant ansatz together with a quantum circuit structure that enforces Gauss’s law. As a proof of principle, we benchmark our simulation protocol on a small lattice system of $2\times2$, demonstrating the identification of density-induced level-crossings in terms of the particle number of the fermion flavors. Classical simulations are used to obtain optimized variational parameters, which are then deployed in inference runs on IBM quantum hardware. We conclude by discussing hardware limitations and prospects for scaling this method to larger systems.
\end{abstract}

\maketitle

\section{Introduction}
Gauge theories play a fundamental role in the description of nature. 
In the Standard Model, non-abelian gauge theories describe the strong interactions (quantum chromodynamics, QCD) and the electroweak interactions, while abelian gauge theories  account for quantum electrodynamics (QED). Among these, simulating QCD poses particular challenges due to its strongly interacting nature. Lattice gauge theory has provided us with various insights into QCD, but regimes such as finite density, the $\theta$-term, and transport properties of the quark-gluon plasma remain difficult to access. Furthermore, real-time dynamics, essential for understanding phenomena such as the out-of-equilibrium evolution after heavy-ion or proton collisions, are inaccessible with conventional lattice QCD techniques.

To address this challenge, there has been growing interest in studying the Hamiltonian formulation of lattice theories as a complementary approach (see, e.g., Refs.~\cite{DiMeglio:2023nsa,Funcke:2023jbq,Magnifico:2024eiy} for recent reviews). However, current methods such as exact diagonalization, tensor networks, and variational quantum algorithms remain limited to small systems due to both hardware constraints and the challenge of discretizing the Hilbert space. As a result, a full Hamiltonian treatment of QCD in (3+1)D is still out of reach. A natural strategy is therefore to incrementally increase the dimensionality of the system, studying lower-dimensional models that preserve key features of QCD. The (1+1)D Schwinger model has long served this role, exhibiting confinement, chiral symmetry breaking, and a mass gap~\cite{Coleman:1976uz}. Recently, it has been simulated with multiple fermion flavors at finite density on superconducting quantum hardware~\cite{Schuster2024} and on a trapped-ion quantum processor~\cite{Melzer2025}.

A natural next step is QED in 2+1D (QED$_3$), which retains key features such as confinement~\cite{Appelquist:1981vg}. Beyond being a testbed for QCD, QED$_3$ is physically rich in its own right. In condensed matter systems, QED$_3$ with multiple fermion flavors and chemical potentials can serve as an effective description of the experimentally elusive quantum spin
liquids~\cite{Hermele2004, Assaad2005spin, Hermele2005, Hermele2007erratum, He2017physref} and cuprate
superconductors~\cite{AffleckMarston1988, Affleck1989, Lee1998superconductor,Rantner2001, Rantner2002superconductor}. This interplay between QED$_3$ and fermionic matter has recently been explored with quantum Monte Carlo methods \cite{Xu:2018wyg,Chen2025}. In addition, quantum spin liquids have also been studied using exact diagonalization~\cite{Wietek:2023qsf}. 
The first quantum simulations of QED$_3$ were recently reported in Refs.~\cite{Crippa2024withEnrico, Meth2023}, focusing on a single fermion flavor at zero density.
QED$_3$ with two fermion flavors at finite density has been simulated using variational Monte Carlo methods \cite{Bender2023}, while QED$_3$ with a single fermion flavor and non-zero total charge has been investigated with neural flow wavefunctions \cite{chen2022} and tensor networks \cite{Felser2020}. Complementary to these matter-focused studies, further recent works on 2+1D QED focusing on the gauge sector, specifically truncation schemes, include~\cite{Miranda-Riaza:2025fus,Mou:2025iiu}, while cold-atom simulator proposals exploiting higher-spin representations have also been put forward~\cite{Majcen:2026sfm}.

In this work, we present the first proof-of-principle demonstration of QED$_3$ with multiple fermion flavors at finite density on quantum hardware. To this end, we employ the variational quantum eigensolver (VQE) \cite{Peruzzo2014}, a resource-efficient alternative to quantum phase estimation~\cite{Kitaev:1995qy} that is well-suited for noisy intermediate-scale quantum (NISQ) devices \cite{Preskill_2018}. For the circuit ansatz, we generalize the QED$_3$ construction of Ref.~\cite{Crippa2024} to incorporate two fermion flavors at finite density. We classically simulate the VQE and perform inference runs on quantum devices, thereby benchmarking our finite-density state-preparation protocol on small lattice sizes of $2\times2$. Our (2+1)D finite-density protocol marks a first step toward real-time evolution of finite-density systems in (2+1)D, relevant for quench dynamics (see, e.g., Ref.~\cite{Cobos:2025krn, Cao:2026qky}). This approach is scalable to larger lattices through advanced adaptive ansätze such as (SC)-ADAPT-VQE \cite{Farrell2024} and $\text{(SC)}^2$-ADAPT-VQE \cite{Gustafson2024}, which we leave for future study.
The paper is structured as follows. In \cref{sec:ModelsAndMethods}, we first introduce the Hamiltonian in \cref{subsec:Hamiltonian} and discuss the phase structure in \cref{subsec:PhaseStructure}. The VQE protocol is detailed in \cref{sec:VQEprotocol}, where \cref{subsec:QC} presents the quantum circuit and \cref{subsec:Optimizer} describes the classical optimizer. The results are given in \cref{sec:Results}, with classical simulations in \cref{subsec:CS} and inference runs in \cref{subsec:InferenceRun}. We conclude with a discussion of the results in \cref{sec:DiscussionAndConclusions}. The source code is available at Ref.~\cite{CrippaEtAl2025_LatticeQED}.

\section{Model and Methods}
\label{sec:ModelsAndMethods}

\subsection{Lattice Hamiltonian}
\label{subsec:Hamiltonian}

The model studied in this work is an extension of the one presented in Ref.~\cite{Crippa2024} for multiple flavors of fermions with different masses and chemical potentials.

Thus, we consider a (2+1)-dimensional lattice formulation of QED with staggered fermions of multiple flavors and compact gauge links. The total Hamiltonian is given by
\begin{equation}
    \hat{H} = \hat{H}_E + \hat{H}_B + \hat{H}_m + \hat{H}_\text{kin} + \hat{H}_\mu.
    \label{eq:H}
\end{equation}
The electric field contribution reads
\begin{equation}
    \hat{H}_E = \frac{g^2}{2} \sum_{\vec{r}} \big( \hat{E}_{\vec{r},x}^2 + \hat{E}_{\vec{r},y}^2 \big).
\end{equation}
Here, $\hat{E}_{\vec{r},\nu}$ is the electric field on the gauge link starting at site $\vec{r}$ and going in $\nu$ direction. The coupling strength is denoted with $g$. As in Ref.~\cite{Crippa2024}, the electric fields are truncated to 
\begin{equation}
    \hat{E}|e\rangle = e|e\rangle,\quad e\in[-l,\dots,l],
\end{equation}
where the integer $l$ is the truncation parameter. The effect of this truncation is discussed in \Cref{app:Trunc}.
The magnetic plaquette term is
\begin{equation}
    \hat{H}_B = -\frac{1}{2a^2 g^2} \sum_{\vec{r}} \big( \hat{P}_{\vec{r}} + \hat{P}^\dagger_{\vec{r}} \big),
\end{equation}
with the plaquette operator defined as
\begin{equation}
    \hat{P}_{\vec{r}} = \hat{U}_{\vec{r},x} \hat{U}_{\vec{r}+x,y} \hat{U}^\dagger_{\vec{r}+y,x} \hat{U}^\dagger_{\vec{r},y},
\end{equation}
where $\hat{U}_{\vec{r},\nu}$ acts as a ladder operator for $\hat{E}_{\vec{r},\nu}$ in its eigenbasis, and $a$ denotes the lattice spacing. For all numerical results, we set $a=1$.
The fermionic sector includes the fermionic mass term,
\begin{equation}
    \hat{H}_m = \sum_{\vec{r},f} (-1)^{r_x + r_y} m_f\, \hat{\phi}^\dagger_{\vec{r},f} \hat{\phi}_{\vec{r},f},
\end{equation}
with $\hat{\phi}_{\vec{r},f}$ being a one-component spinor at site $\vec{r}$ of flavor $f$ and mass $m_f$.
Furthermore, there is the kinetic hopping term,
\begin{equation}
\begin{split}
    \hat{H}_\text{kin} &= \frac{i}{2a} \sum_{\vec{r},f} \Big( \hat{\phi}^\dagger_{\vec{r},f} \hat{U}_{\vec{r},x} \hat{\phi}_{\vec{r}+x,f} - \text{h.c.} \Big) \\
    &\quad - \frac{1}{2a} \sum_{\vec{r},f} (-1)^{r_x + r_y}\Big( \hat{\phi}^\dagger_{\vec{r},f} \hat{U}_{\vec{r},y} \hat{\phi}_{\vec{r}+y,f} + \text{h.c.} \Big).
\end{split}
\end{equation}
Finally, the chemical potential contribution reads
\begin{equation}
    \hat{H}_\mu = \sum_{\vec{r},f} \mu_f\, \hat{\phi}^\dagger_{\vec{r},f} \hat{\phi}_{\vec{r},f},
\end{equation}
with $\mu_f$ being the chemical potential of flavor $f$.
Note that our sign convention for the chemical potential differs from the standard grand-canonical form ($H - \mu N$), following instead the convention of Ref.~\cite{Bender2023}. This does not affect the physical content of our analysis: we work at fixed total particle number $N_1 + N_2$, and our interest is solely in the redistribution of particles between the two flavors as the relative chemical potential $\mu_1 - \mu_2$ is varied. A global sign redefinition of both chemical potentials leaves this quantity unchanged, and therefore does not affect any physically relevant observable or intuition in our study.
Turning to the gauge sector, physical states $|\Phi\rangle$ must satisfy Gauss’ law at each lattice site:
\begin{equation}
    \Bigg[ \sum_{\mu=x,y} \big( \hat{E}_{\vec{r},\mu} - \hat{E}_{\vec{r}-\mu,\mu} \big) - \sum_f \hat{q}_{\vec{r},f} \Bigg] |\Phi\rangle = 0,
\end{equation}
where the local dynamical charge is
\begin{equation}
    \hat{q}_{\vec{r},f} = \hat{\phi}^\dagger_{\vec{r},f} \hat{\phi}_{\vec{r},f} - \frac{1}{2} \big[1 + (-1)^{r_x + r_y + 1}\big].
\end{equation}
Following Ref.~\cite{Crippa2024}, we restrict our simulation to the gauge-invariant subspace by explicitly encoding Gauss’s law into the ansatz. This means that the equations are explicitly solved, such that some gauge links can be expressed in terms of others. 
In the case of a $2\times2$ lattice, Gauss' law corresponds to four individual equations. With open boundary conditions (OBC), there are four links and four charges. Solving the equations then eliminates three dynamical links and sets the total charge to zero.
Consequently, for this system, only one instead of four gauge links is unconstrained. This link is simulated, while the other gauge links are expressed in terms of this link. In the following, we elaborate on this minimal setup, as it provides a substantial reduction of the gauge-field Hilbert space.

Despite this simplification, which results in simulating just a single link, the matter sector retains a genuine two-dimensional structure. In particular, the fermionic Hamiltonian includes hopping terms along both lattice directions, such that the fermions still probe the full $2$D connectivity of the lattice. Furthermore, the gauge sector remains nontrivial at the level of plaquette interactions. After imposing Gauss’s law on the $2\times2$ lattice with open boundary conditions, the plaquette operator $\hat{P}_{\vec{r}} + \hat{P}^{\dagger}_{\vec{r}}$ reduces to an off-diagonal operator acting on the remaining dynamical link. In the electric-flux basis, this corresponds to a raising and lowering operator, thereby coupling different flux sectors.

To assess the extent to which the observed density-induced level crossings are influenced by the minimal $2\times2$ OBC geometry, we additionally perform exact-diagonalization (ED) calculations for systems with a less constrained gauge structure, including a $2\times2$ lattice with periodic boundary conditions (PBC) as well as $2\times3$ and $2\times4$ lattices with OBC (see Appendix~C). For the same gauge coupling $g=5$, we find that the qualitative level-crossing structure persists across all geometries considered. This indicates that the observed behavior is not an artifact of the specific Gauss-law reduction in the $2\times2$ OBC system.

\subsection{Phase Structure}
\label{subsec:PhaseStructure}

The zero-temperature phase diagram of two massless fermions at finite density can be analyzed through the particle number of the fermion flavors, following the approach of the (1+1)D study in Refs.~\cite{Schuster2024, Banuls:2016gid}, but now applied to a (2+1)D theory. For this analysis, we re-write the flavor chemical potentials in terms of the chemical potential coupled to the total charge and in terms of the charge difference (isospin) of the two species (see Ref.~\cite{Narayanan:2012qf} and \Cref{eq:rewrite} below). 

In (1+1)D, the partition function and therefore the phase diagram is independent of the sum of the two chemical potentials, therefore, the net charge can be set to zero~\cite{Narayanan:2012qf}. 
Indeed, Ref.~\cite{Narayanan:2012qf} set the net charge to zero and obtained the density-induced level-crossing points in the sign-problem free region, in terms of the isospin chemical potential. References~\cite{Banuls:2016gid} and~\cite{Schuster2024} went beyond and studied the same phase diagram in the sign-problem-afflicted regime of non-zero net chemical potential, using tensor networks and quantum computing, respectively.

\begin{figure}
    \centering
    \includegraphics[width=\linewidth]{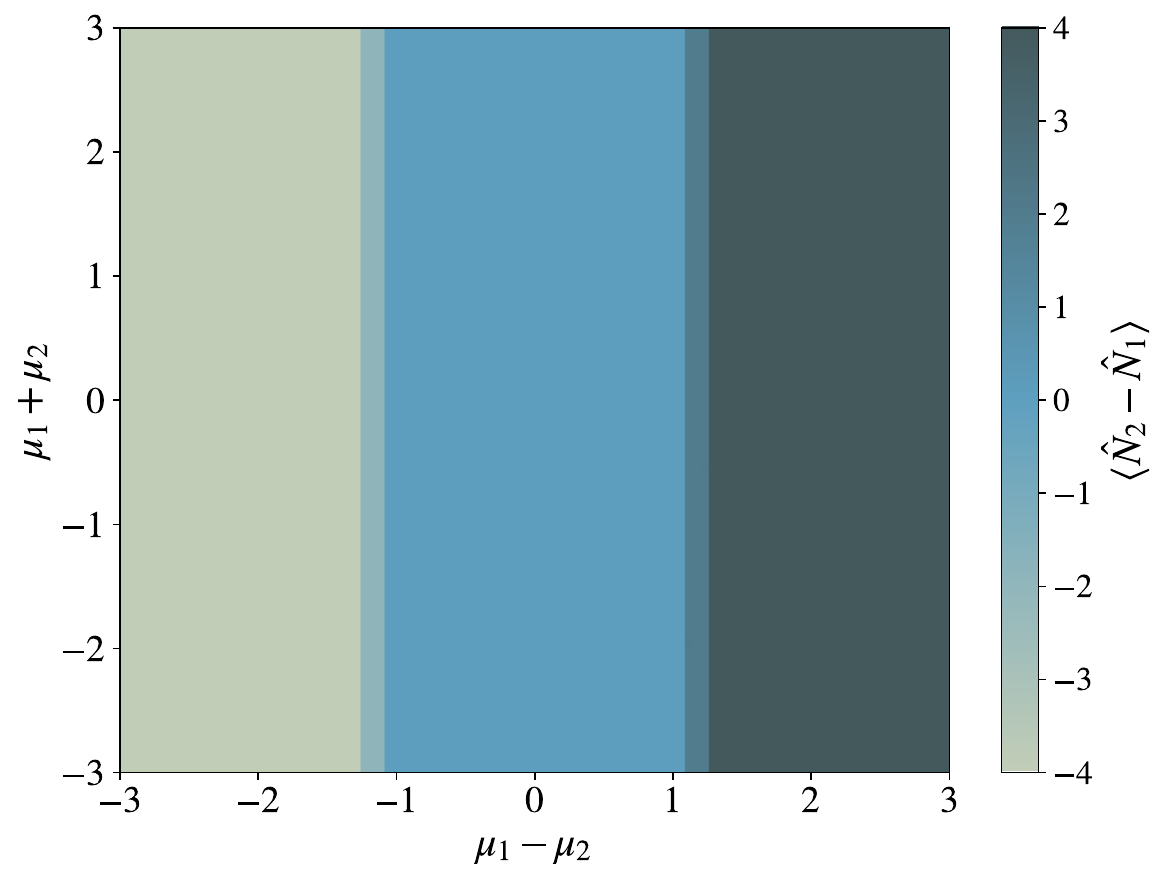}
    \caption{Difference in particle number $\langle\hat{N}_2-\hat{N}_1\rangle$ depending on the sum and difference of $\mu_1$ and $\mu_2$ for a $2\times2$ system with open boundary conditions at $g=2$ with two flavors of massless fermions.}
    \label{fig:VaryingBoth}
\end{figure}

In (2+1)D, no theoretical predictions exist but many considerations parallel those in lower dimensions. Indeed, also in (2+1)D, the phase diagram is independent of the sum of the two chemical potentials, as we show in \cref{fig:VaryingBoth} using exact diagonalization. 

For simulating the sign-problem-afflicted regime of non-zero net chemical potential in our work, we now derive a formula for the locations of the density-induced level-crossings in (2+1)D, following the analogous derivation in (1+1)D from Ref.~\cite{Schuster2024}. 
 
For this, we first define the total number operator for each of the $F$ flavors,
\begin{equation}
    \hat{N}_f = \sum_{\vec{r}} \hat{\phi}^\dagger_{\vec{r},f} \hat{\phi}_{\vec{r},f},
\end{equation}
and note that $[\hat{H}, \hat{N}_f] = 0$. This enables us to express the ground-state energy in each phase sector as
\begin{equation}
    E_{(N_f)}(\mu_f) = \sum_{f=1}^{F} \mu_f N_f + E_\text{ind},
\end{equation}
where $E_\text{ind}$ is a constant for each sector and $N_f$ is the eigenvalue of the operator $\hat{N}_f$. From now on, we only consider two fermion flavors \footnote{However, in \cref{app:ThreeFermions} a short discussion of the case including three fermion flavors can be found.}. Thus, we define
\begin{equation}
\label{eq:rewrite}
      \mu_1 = 0, \quad \mu_2 = \mu, \quad N = N_1 + N_2, \quad \Delta N = N_2 - N_1,  
\end{equation}
yielding the ground-state energy
\begin{equation}
    E_{(N_1,N_2)}(\mu) = \frac{1}{2} \mu N + \frac{1}{2} \mu \Delta N + E_\text{min}.
\end{equation}
The density-induced level-crossing occurs at the critical chemical potential $\mu_\text{LC}$, where energies in different sectors become degenerate, i.e.,
\begin{equation}
    E_{(N_1,N_2)}(\mu_\text{LC})=E_{(\tilde{N}_1,\tilde{N}_2)}(\mu_\text{LC})\,.
\end{equation}
Here, the second sector is denoted by a tilde. 
This equation can then be solved for
\begin{equation}
    \mu_\text{LC} = 2\, \frac{E_\text{min}(N_1,N_2) - E_\text{min}(\tilde{N}_1,\tilde{N}_2)}
    {\Delta N(\tilde{N}_1,\tilde{N}_2) - \Delta N(N_1,N_2)}.
\end{equation}
To determine the constant values of each phase, arbitrary points $\mu^*$ and $\tilde{\mu}^*$ in both phases can be used. This way, the final expression for the critical point reads
\begin{equation}
\label{eq:PTeq}
    \mu_\text{LC} = 2\cdot\frac{E_{(N_1,N_2)}(\mu^*)-E_{(\tilde{N}_1,\tilde{N}_2)}(\tilde{\mu}^*)-N_2\mu^*+\tilde{N}_2\tilde{\mu}^*}{\Delta N(\tilde{N}_1,\tilde{N}_2)-\Delta N(N_1,N_2)}\,.
\end{equation}
This expression will be used later on to determine the location of the density-induced level-crossing from the data obtained on quantum hardware.  

\section{VQE Protocol}
\label{sec:VQEprotocol}

The VQE needs a parametrized ansatz that is able to capture the ground state wavefunction as a quantum circuit. Moreover, an optimizer is required to find the parameters minimizing the energy. Both components will be presented in the following.

\subsection{Quantum Circuit Ansatz}
\label{subsec:QC}

The variational ansatz comprises three components: (i) fermion state encoding, (ii) gauge link representation, and (iii) entangling gates ensuring the matter-gauge coupling. The schematic structure is shown in \cref{fig:CircuitStructure}. Due to limitations on the available hardware, this work focuses on a $2\times2$ system with OBC. As discussed above, this means that three out of four links can be fixed by Gauss's law and a total of eight fermions are present. However, the ansatz can be extended to larger systems or PBC as well.

\begin{figure}
    \centering
    \begin{quantikz}
        \lstick{$|0\rangle$} & & \gate[3]{\text{Gauge}} & \gate[6]{\text{Entanglement}} &\\
        \setwiretype{n} \lstick{} & \myvdots & &\\
        \lstick{$|0\rangle$} & & & &\\
        \lstick{$|0\rangle$} & & \gate[3]{\text{Fermion}} & &\\
        \setwiretype{n} \lstick{} & \myvdots & &\\
        \lstick{$|0\rangle$} & & & &
    \end{quantikz}
    \caption{Schematic structure of the variational ansatz combining gauge, fermion, and entanglement subcircuits.}
    \label{fig:CircuitStructure}
\end{figure}

\begin{figure}
    \centering
    \includegraphics[width=0.8\linewidth]{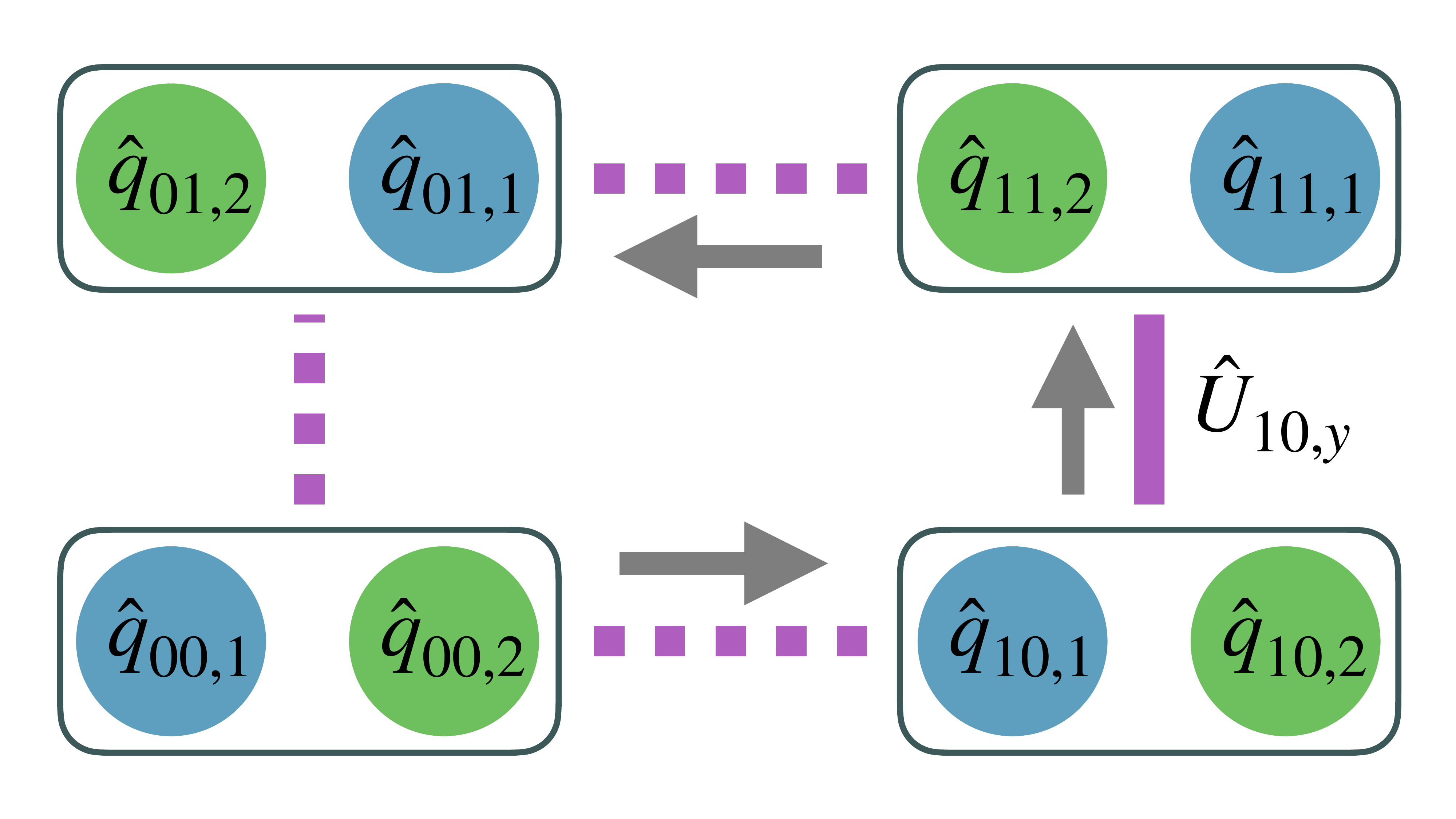}
    \caption{The $2\times2$ system with open boundary conditions. Dotted purple lines represent fixed gauge links, while the solid purple line depicts a free gauge link. At each physical lattice site, two flavors of fermions are present and thus it is mapped to two qubits. The gray arrows indicate the ordering of the fermionic degrees of freedom in the Jordan-Wigner transformation.}
    \label{fig:LatticeDrawing}
\end{figure}

The mapping of the lattice to a string of qubits can be performed as indicated by \cref{fig:LatticeDrawing}. By applying Jordan-Wigner mapping \cite{Jordan1928, Verstraete_2005} the fermionic degrees of freedom can be mapped to qubits using the ordering indicated by the arrows. Each lattice site includes two flavors of fermions and thus is mapped to two qubits. Therefore, the simulated lattice is effectively of double the width of the physical lattice. 

For the gauge sector, we adopt the Gray encoding at truncation $l=1$ from Ref.~\cite{Crippa2024} (see \cref{fig:GaugeCircuit}). Since $l=1$ requires three states ($\ket{-1},\ket{0}$ and $\ket{1}$ \footnote{These states refer to the physical states to mark a distinction from the encoded ones.}), but any number of qubits results in a number of basis states being a power of two, one basis state needs to be excluded. This encoding on two qubits removes the state $|01\rangle$, as the state of the second qubit can only be changed when the first is in state $|1\rangle$. An important point to note here is that hardware noise can cause violations of Gauss law causing the excluded state $|01\rangle$ to be reached with a small but non-zero probability. We have quantified this effect in appendix~\ref{app:Leakage}.

\begin{figure}
    \centering
    \begin{quantikz}
        \lstick{$U_{10,y,0}\to|0\rangle$} & \gate{R_y(\theta_0)} & \ctrl{1} &\\
        \lstick{$U_{10,y,1}\to|0\rangle$} & & \gate{R_y(\theta_1)} &
    \end{quantikz}
    \caption{Gauge field encoding circuit using Gray code at truncation $l=1$, following Ref.~\cite{Crippa2024}.}
    \label{fig:GaugeCircuit}
\end{figure}

The fermionic subcircuit generalizes the structure in Ref.~\cite{Schuster2024} to two flavors, with additional inter-flavor mixing layers (see \cref{fig:FermionCircuit}). Here, the $U^{xy}(\theta)$ gates are defined as 

\begin{equation}
\label{eq:UxyGate}
U^{xy}_{ij}(\theta) = e^{-i\frac{\theta}{2}\cdot(X_iX_j+Y_iY_j)}\,.    
\end{equation}
These gates ensure that only states with the same number of qubits in state $|0\rangle$ and $|1\rangle$ can be reached. The structure of the circuit in Ref.~\cite{Schuster2024} can be seen shaded as gray and white for each flavor in \cref{fig:FermionCircuit}. The inter-flavor mixing layer can be seen in light blue. The last layer is made up of $R_Z(\theta)$ gates for each qubit to correct the phases. This layer is not shown for better visibility.

\begin{figure*}
    \centering
    \begin{quantikz}
        \lstick{$q_{00,1}\to|0\rangle$} & & \colorgate{3}{gray!30}{U^{xy}\left(\theta_0\right)}{dash dot} & &  & & \colorgate{2}{blue!20}{U^{xy}\left(\theta_6\right)}{dotted} & \\
        \lstick{$q_{00,2}\to|0\rangle$} & & & \gate[3]{U^{xy}\left(\theta_2\right)} & & & & \\
        \lstick{$q_{10,1}\to|0\rangle$} & \gate{X} & & & \colorgate{3}{gray!30}{U^{xy}\left(\theta_4\right)}{dash dot} & & \colorgate{2}{blue!20}{U^{xy}\left(\theta_7\right)}{dotted} &\\
        \lstick{$q_{10,2}\to|0\rangle$} & \gate{X} & & & & \gate[3]{U^{xy}\left(\theta_5\right)} & & \\
        \lstick{$q_{11,1}\to|0\rangle$} & & \colorgate{3}{gray!30}{U^{xy}\left(\theta_1\right)}{dash dot} & & & & \colorgate{2}{blue!20}{U^{xy}\left(\theta_8\right)}{dotted} & \\
        \lstick{$q_{11,2}\to|0\rangle$} &  & & \gate[3]{U^{xy}\left(\theta_3\right)} & & & & \\
        \lstick{$q_{01,1}\to|0\rangle$} & \gate{X} & & & & & \colorgate{2}{blue!20}{U^{xy}\left(\theta_9\right)}{dotted} & \\
        \lstick{$q_{01,2}\to|0\rangle$} & \gate{X} & & & & & & \\
    \end{quantikz}
    \caption{Fermion encoding circuit for a $2\times2$ system with two fermion flavors. The different parts are highlighted by their color. The cascade structure is shown in light gray and white (dash dotted and solid), while the inter-flavor gates are shown in light blue (dotted). The different line style and color is only for highlighting and the gates are defined in the same way.}
    \label{fig:FermionCircuit}
\end{figure*}

Controlled $U^{xy}$-gates entangle the gauge and matter sectors while maintaining gauge invariance as shown in~\cite{Crippa2024withEnrico}.

\subsection{Classical Optimization}
\label{subsec:Optimizer}

The ansatz contains 24 variational parameters optimized using the COBYLA algorithm~\cite{Powell1994}. The fermionic circuit shown in \cref{fig:FermionCircuit} contains 10 parameters, while the gauge circuit in \cref{fig:GaugeCircuit} has two variational parameters. The additional $R_Z$-layer adds another 8 parameters. The remaining four parameters are hidden in the entanglement circuit shown in \cref{fig:CircuitStructure}. While VQE-specific optimizers such as NFT~\cite{NFT2020}, can offer faster convergence, their additional requirements significantly increase the circuit depth in our case. Specifically, for NFT, all gates of the form 
\begin{equation}
    R_j(\theta_j) = \text{exp}\!\left(-\frac{i\theta_j}{2}A_j\right)\,,
\end{equation}
need to fulfill the requirement $A_j^2=\mathbb{I}$. However, this is not the case for the gates considered in this work. The generator $A_j$ for $U^{xy}_{ij}(\theta)$ is $X_i X_j + Y_i Y_j$. Squaring this operator yields
\begin{equation}
(X_i X_j + Y_i Y_j)^2 = 2\mathbb{I} - 2 Z_i Z_j,.
\end{equation}
Consequently, the $U^{xy}_{ij}(\theta)$ gates in \cref{eq:UxyGate} satisfy $A_j^2 \neq \mathbb{I}$. To make NFT applicable, one would need $A_j^2 = \mathbb{I}$, which in turn would require adding an additional $Z_i Z_j$ term to the exponent in \cref{eq:UxyGate}, thereby increasing the circuit depth. Consequently, COBYLA is employed with a maximum of 1,000 iterations, which ensures convergence. To obtain the optimal parameters, the simulation of the quantum computer is performed in the limit of infinite measurements. 

\subsection{Scaling of the Ansatz Circuit}
\label{app:ScalingAnsatz}

In this work a $2\times2$ system with open boundary conditions is considered. However, the ansatz circuit can also be extended to different system sizes and boundary conditions. In this section, a few different systems will be discussed in terms of qubits and depth of the circuit. For this purpose \cref{tab:ScalingAnsatzCircuit} shows the properties of the ansatz circuits for $n_\text{flavors}=2$ and $n_\text{flavors}=1$, respectively. It can be seen that going from periodic to open boundary conditions reduces the circuit depth by a factor of $\mathcal{O}(10)$, while the circuit depth only moderately increases when going from $N_f=1$ to $N_f=2$ flavors. This is because increasing the number of flavors only affects the fermionic sector, while switching from open to periodic boundary conditions also affects the gauge sector. 

\begin{table}[h]
    \centering
    \begin{tabular}{ccccc}
        \hline
        System & $N_f$ & Depth & \#qubits & \#parameters \\
        \hline
        $2\times2$ OBC & 2 & 69 & 10 & 24 \\
        $2\times2$ PBC & 2 & 644 & 18 & 88 \\
        $2\times2$ OBC & 1 & 26 & 6 & 11 \\
        $2\times2$ PBC & 1 & 601 & 14 & 67 \\
        $2\times4$ OBC & 1 & 211 & 14 & 42 \\
        \hline
    \end{tabular}
    \caption{Ansatz circuit properties for different system sizes at $n_\text{flavors}=2$. Here, \textit{Depth} refers to the layers of multi-qubit gates. The compilation of the circuits is performed for the device \texttt{ibm\_marrakesh} that is introduced in \cref{subsec:InferenceRun}.}
    \label{tab:ScalingAnsatzCircuit}
\end{table}

An analytic description of the scaling of some parts can be given. For simplicity we restrict this to $N_f =2$. Then, the number of qubits for the fermionic part is given as $n_{q,f} = 2\cdot L_x\cdot L_y$. The number of gauge links is $n_{g} = L_x\cdot L_y + 1 +  (-L_x-L_y)$ for PBC (OBC). Thus, the number of qubits for the gauge links is $n_{q,p} = \ceil*{\log_2(2l+1)}\cdot n_g$. Moreover, the number of $U^{xy}$-gates in the fermionic part scales as
\begin{align}
    n\left(U^{xy}\right) &= \frac{n_{q,f}}{4}\left( \frac{n_{q,f}}{2} + 1 \right) + L_x\cdot L_y \\
    &= \frac{1}{2}\left( L_x \cdot L_y \right)^2 + \frac{3}{2}L_x\cdot L_y\,.
\end{align}
The depth is $d(U^{xy}) = L_x\cdot L_y+1$. The depth of the other circuits cannot be expressed in closed form, as the circuit encoding a gauge link changes depending on the truncation.

We emphasize that the ansatz can, in principle, be efficiently scaled to larger lattice sizes by employing advanced adaptive methods such as (SC)-ADAPT-VQE \cite{Farrell2024} and $\text{(SC)}^2$-ADAPT-VQE \cite{Gustafson2024}. A detailed investigation of this scaling is left to future work.

\section{Results}
\label{sec:Results}

\subsection{Classical Simulation}
\label{subsec:CS}

To benchmark the protocol, we perform noiseless classical simulations for a $2\times2$ lattice with open boundary conditions, gauge coupling $g=5$, and truncation $l=1$. In this large coupling regime, the support of the electric field remains small and this truncation suffices. The effect of this truncation in our setup is discussed in \Cref{app:Trunc}.
In principle, various combinations of these parameters can be studied. However, smaller couplings lead to a larger entanglement which then requires a deeper circuit. 
 
\begin{figure}
    \centering
    \includegraphics[width=\linewidth]{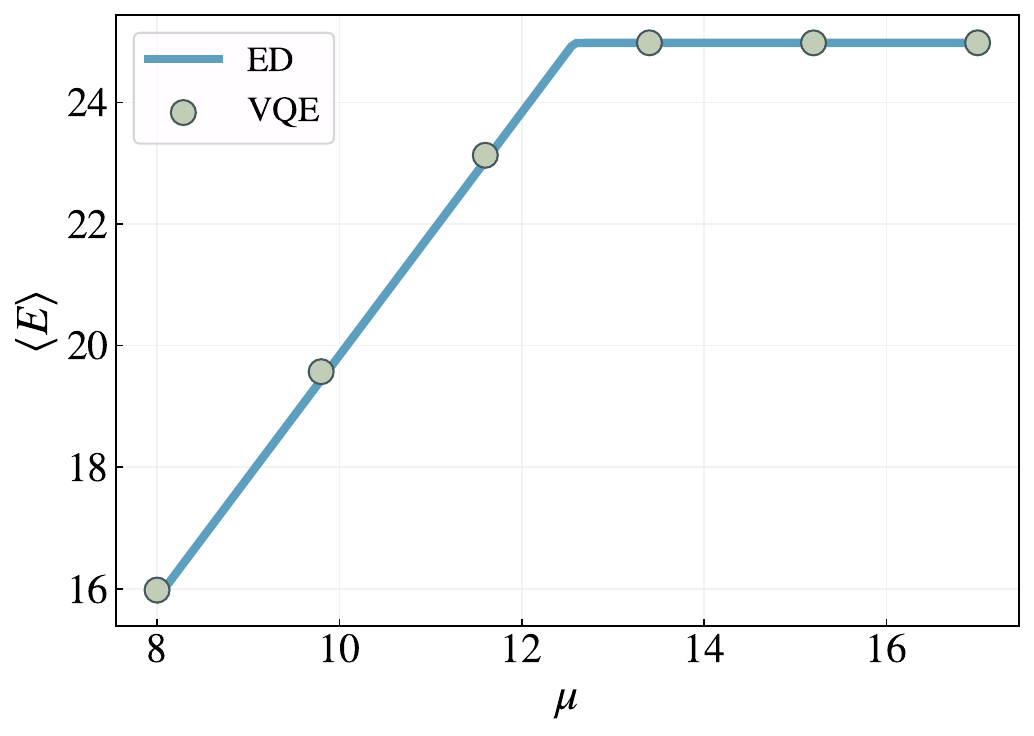}
    \caption{Ground-state energies $\langle E\rangle$ against the chemical potential $\mu$ from the classical VQE simulation for a $2\times2$ system. As these are calculations with infinite shots, there are no error bars.}
    \label{fig:ClassSimEnergy}
\end{figure}

\begin{figure}
    \centering
    \includegraphics[width=\linewidth]{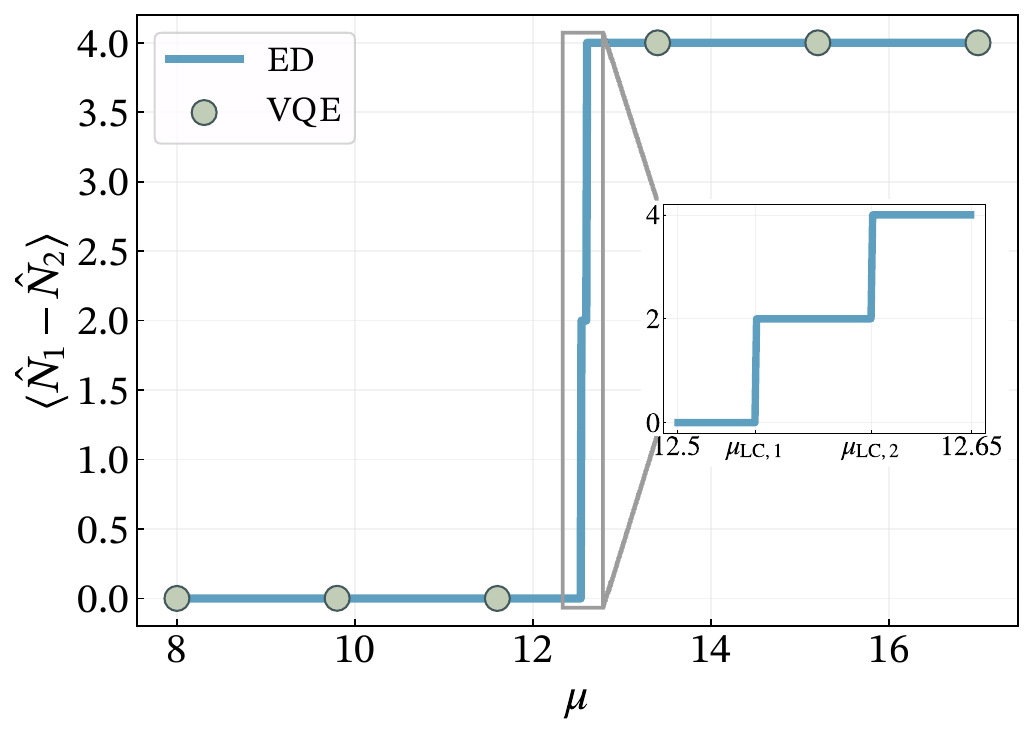}
    \caption{Difference in fermion number $\Delta N$ against the chemical potential $\mu$ from the classical VQE simulation. As these are calculations with infinite shots, there are no error bars.}
    \label{fig:ClassSimNum}
\end{figure}

Therefore, we focus on the large-coupling region of $g=5$.
The resulting ground-state energies and particle number differences are shown in \cref{fig:ClassSimEnergy,fig:ClassSimNum}. The emergence of level crossings is clearly visible, similar to~\cite{Schuster2024}. A small intermediate phase appears near the transition region, which cannot be well resolved with the VQE. The width of this region increases with decreasing coupling strength. An analysis of this region can be found in \cref{app:IR}.

The difference in particle numbers $N_1$ and $N_2$ shows the expected steps of height $\Delta N_\text{first region} - \Delta N_\text{second region} = 2$ as we consider two flavors of fermions and thus each increase in $N_1$ corresponds to a decrease of $N_2$ and vice versa.

Due to computational limitations, we restrict our study to a $2\times2$ lattice. While the positions of these level crossings are expected to shift with increasing system size, their qualitative nature and magnitude, $\Delta N_\text{first region} - \Delta N_\text{second region} = 2$, remain unaffected.

\subsection{Analysis of the Intermediate Region}
\label{app:IR}

\begin{figure}
    \centering
    \includegraphics[width=\linewidth]{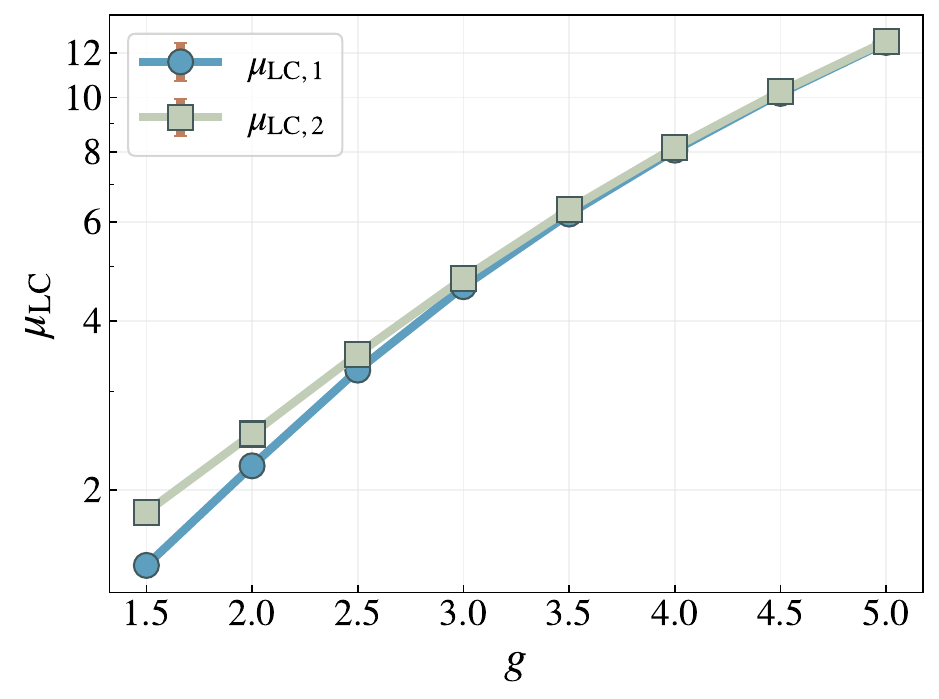}
    \caption{Critical values of the chemical potential $\mu_{\rm pt}$, as computed in \cref{subsec:CS} using ED, as a function of the gauge coupling $g$. Here, $\mu_{\rm LC, 1}$ and $\mu_{\rm LC,2}$ correspond to the edges of the intermediate region characterized by $\langle \hat{N}_1-\hat{N}_2 \rangle =2$, as shown for $g=5$ in \Cref{fig:ClassSimNum}. Note that the intermediate region has a decreasing width and an increasing center with increasing $g$. The uncertainties are smaller than the markers.}
    \label{fig:IRanalysis}
\end{figure}

\begin{figure}
    \centering
    \includegraphics[width=\linewidth]{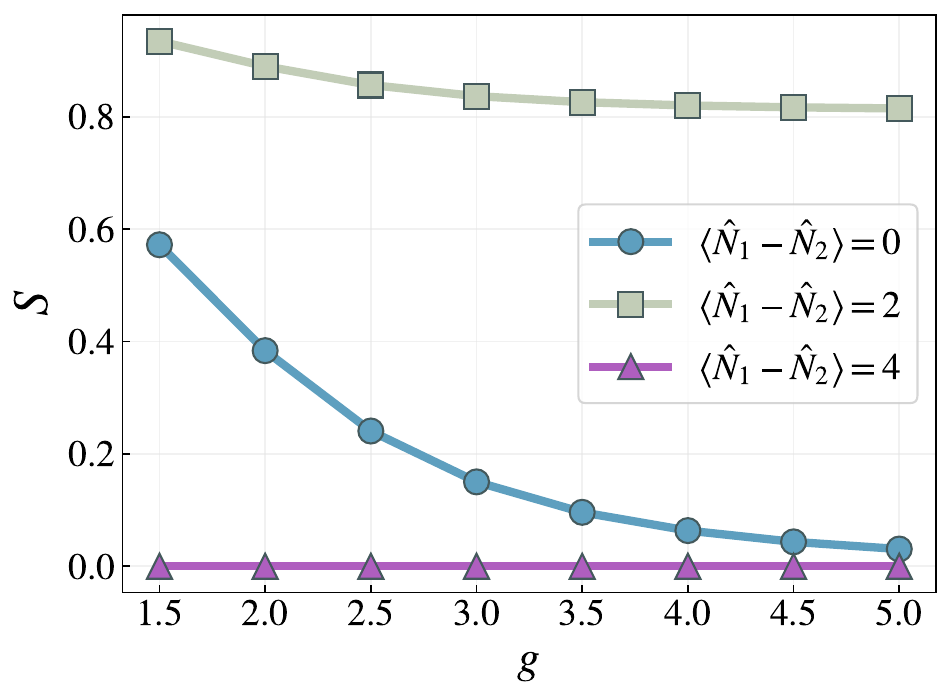}
    \caption{Von Neumann entropy $S$ between the gauge link and the fermion system for different coupling strengths $g$ in all three regions. The ground state was calculated with ED in each of the three particle number sectors and the von Neumann entropy was obtained using \cref{eq:vonNeumannEntropy}.}
    \label{fig:IRentanglement}
\end{figure}

In \cref{fig:ClassSimNum}, the intermediate region with $\langle\hat{N}_1-\hat{N}_2\rangle=2$ can be seen, although it appears to be of a small width. If the coupling is modified, the size of the intermediate region changes. This can be seen in \cref{fig:IRanalysis} where the positions of both level crossings are shown. The difference of these decreases with $g$, making the region difficult to resolve. Additionally, as $g$ increases, the center of the intermediate region shifts noticeably towards larger values. This would naturally make it an obvious choice to simulate $g$ as small as possible. However, in order to keep the circuit depth as small as possible, the entanglement in the system should be as small as possible, as this allows for fewer layers. Thus, in \cref{fig:IRentanglement}, the von Neumann entropy between the gauge link and the fermion system
\begin{equation}
\label{eq:vonNeumannEntropy}
    S = -\Tr{\rho_G\log_2{\rho_G}}\quad\text{with}\quad \rho_G=\Tr_F\{\rho\}
\end{equation}
is calculated for the ground state. Here, $\rho$ is the density matrix describing the full subsystem and the subscripts $G$ and $F$ denote the gauge and fermion subsystem, respectively. Consequently, $S=0$ mean completely decoupled fermion and gauge systems.

In \cref{fig:IRentanglement}, it can be seen that the intermediate region exhibits a comparatively large entanglement regardless of the coupling, while the entanglement entropy in the other regions approaches $S=0$ for increasing $g$. Therefore, investigating a coupling at which the region with high entanglement cannot be resolved is favorable for simulation on real hardware.

\subsection{Inference Run on Quantum Hardware}
\label{subsec:InferenceRun}

The optimized parameters obtained from classical simulation are deployed for inference runs on IBM quantum hardware. Due to noise and limited circuit depth, practical results highlight the challenges in preserving gauge symmetry and capturing the full phase structure with current devices.

\begin{figure}
    \centering
    \includegraphics[width=\linewidth]{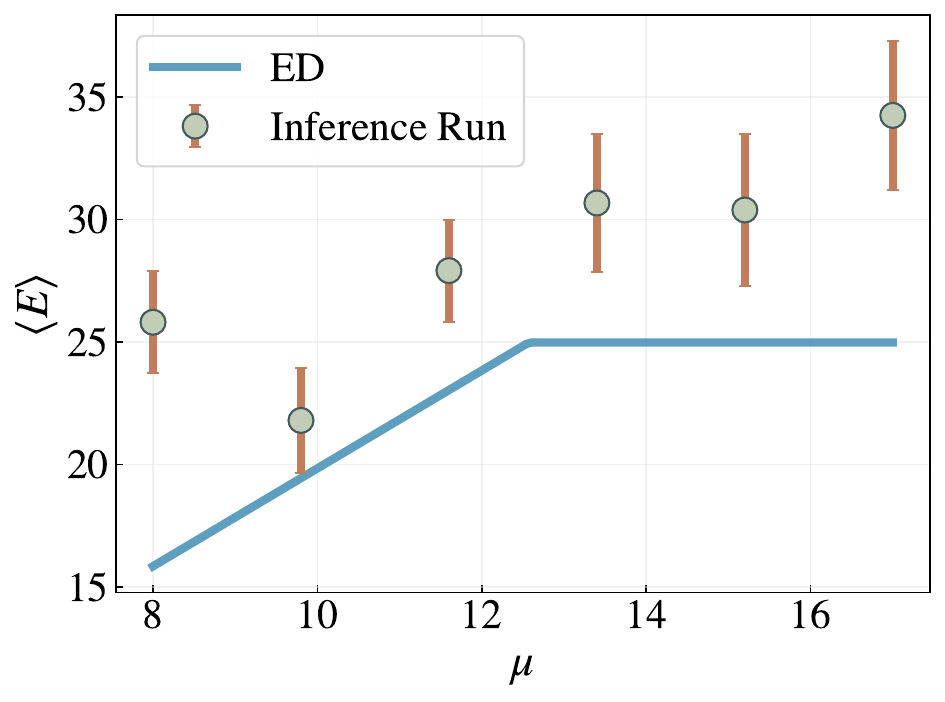}
    \caption{The energy $\langle E\rangle$ as a function of the chemical potential $\mu$ measured on \texttt{ibm\_marrakesh} with $n_\text{shots}=8,\!192$ after applying ZNE for the system described above.}
    \label{fig:InferenceEnergy}
\end{figure}

\begin{figure}
    \centering
    \includegraphics[width=\linewidth]{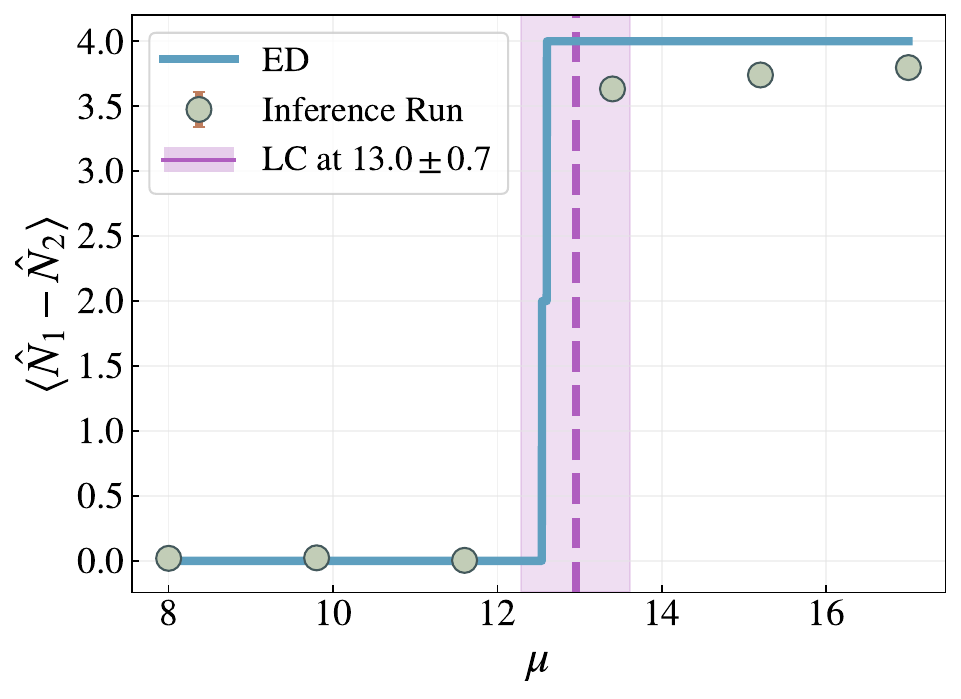}
    \caption{$\hat{N}_1$ and $\hat{N}_2$ as a function of the chemical potential $\mu$ measured on \texttt{ibm\_marrakesh} with $n_\text{shots}=8,\!192$ after applying ZNE for the system described above. The level crossing (LC) is determined with \cref{eq:PTeq} from the data shown in \cref{fig:InferenceEnergy} and here.}
    \label{fig:PhaseTransition}
\end{figure}

For the inference run, the device \texttt{ibm\_marrakesh} is used with $n_\text{shots}=8,\!192$ shots for each data point. The three observables $E$, $N_1$ and $N_2$ are measured and the results are shown in \cref{fig:InferenceEnergy,fig:PhaseTransition}, respectively. All data shown include a Zero-Noise-Extrapolation (ZNE) \cite{Temme2017}. This has been done by applying a linear fit to different noise levels ($\lambda=[1, 1.2, 1.5]$), which have been achieved using gate folding.  Further details are shown in \cref{app:ZNEexplicit}.

One notable observation is that hardware noise has a smaller effect on the number operator than on the Hamiltonian. This can be understood in terms of the condition number, which is lower for the number operator. Further details are provided in \cref{app:NoiseCond}.

Subsequently, the data from this inference run are then used to calculate the critical point of the level crossing (LC) using \cref{eq:PTeq}. The result can be seen in \cref{fig:PhaseTransition}. Both critical points can only be determined using ED, which means no exact values are present. The first is at $\mu=12.540\pm0.001$ and the second at $\mu=12.599\pm0.001$. Thus, both values agree with the critical point determined by the inference run data of $13\pm0.7$.

To resolve both level crossings, data points in the intermediate region have to be taken, which is currently not possible due to limitations on the circuit depth. Consequently, even with advanced error mitigation techniques or a larger number of measurements, the uncertainty would be reduced, but the result would still correspond to an average of the two transition points.

We note that no analytical or numerical results are available for the level crossing points in (2+1)-dimensional QED, neither at finite nor at infinite volume, for the parameter regime studied in this work. The location of the critical points is expected to shift to larger values with increasing system size due to finite-size effects, as observed in the (1+1)D case~\cite{Schuster2024}. For the (2+1)D case, classical simulations of QED at finite density based on Variational Monte Carlo methods~\cite{Bender2023} show that $8\times 8$ lattices are sufficient to avoid finite-size effects. However, a direct comparison of our results to their findings is not possible as they studied a different parameter regime unavailable to us. Therefore, the shift of the location of the transition points due to finite-size effects remain an open question for the parameters chosen in our study.

\subsection{Noise Sensitivity of Operators}
\label{app:NoiseCond}

In \cref{subsec:InferenceRun} it can be seen that the expectation values of the energy are noticeably affected by the hardware noise, while the number operator gives results close to the expected values. To give a quantifiable reason for this, one can define a noise sensitivity analogous to a condition number, which measures how much an arbitrary perturbation can change the measured expectation value.

For an observable $O$ with operator norm $\|O\|_\infty$, the \emph{state-sensitivity} quantifies the worst-case amplification of state-preparation errors, where the error is measured in trace distance $T(\rho,\tilde\rho) = \tfrac12\|\rho - \tilde\rho\|_1$ \cite{Coles2019}:
\begin{equation}
    \kappa_{\mathrm{state}}(O) = 2\,\|O\|_\infty
    = \sup_{T(\rho,\tilde\rho)=1} \big| \operatorname{Tr}[O(\rho - \tilde\rho)] \big| \,.
\end{equation}
This gives the bound
\begin{equation}
    \big| \langle O \rangle_{\tilde\rho} - \langle O \rangle_\rho \big|
    \;\le\; \kappa_{\mathrm{state}}(O)\,T(\rho,\tilde\rho)\,.
\end{equation}

Similarly, the \emph{channel-sensitivity} measures the worst-case amplification of hardware (channel) errors, quantified by the diamond norm $\|\Lambda - \Lambda_0\|_\diamond$ \cite{Benenti2010} between the implemented channel $\Lambda$ and the ideal channel $\Lambda_0$ \cite{Hashim2023}:
\begin{equation}
    \kappa_{\mathrm{chan}}(O) = \|O\|_\infty
    = \sup_{\|\Lambda - \Lambda_0\|_\diamond = 1} 
      \big| \langle O \rangle_{\Lambda} - \langle O \rangle_{\Lambda_0} \big|\,,
\end{equation}
which gives the bound
\begin{equation}
    \big| \langle O \rangle_{\Lambda} - \langle O \rangle_{\Lambda_0} \big|
    \;\le\; \kappa_{\mathrm{chan}}(O)\,\|\Lambda - \Lambda_0\|_\diamond\,.
\end{equation}

For the Hamiltonian $\hat{H}$ in our problem, the operator norm varies with the chemical potential $\mu$. Over the interval
$\mu\in[8,17]$ we find
\begin{equation}
    2174 \;\leq\; \|\hat{H}\|_\infty \;\leq\; 2201\,,
\end{equation}
yielding a state condition number
\begin{equation}
    4348 \;\leq\; \kappa_{\mathrm{state}}(\hat{H}) \;\leq\; 4402\,,
\end{equation}
and
\begin{equation}
    2174 \;\leq\; \kappa_{\mathrm{chan}}(\hat{H}) \;\leq\;2201\,.
\end{equation}
For the number operator $\hat{N}_i$, the operator norm is independent of $\mu$ and takes the value
\begin{equation}
    \| \hat{N}_i \|_\infty = 4\,,
\end{equation}
leading to
\begin{equation}
    \kappa_{\mathrm{state}}(\hat{N}_i) = 8
    \quad\text{and}\quad
    \kappa_{\mathrm{chan}}(\hat{N}_i) = 4\,.
\end{equation}
These values show that small deviations have a much greater impact on the result for the Hamiltonian than for the number operator compliant with our experimental data.

\FloatBarrier

\section{Conclusions}
\label{sec:DiscussionAndConclusions}

In this paper, we presented the first proof-of-principle implementation of QED$_3$ with multiple fermion flavors at finite density on quantum hardware. To this end, we developed a VQE protocol for lattice QED$_3$ with two fermion flavors, enforcing gauge invariance directly at the circuit level. Furthermore, we showed the emergence of density-induced level-crossings in the particle number operator as a function of the chemical potential. To this end, we identified three distinct regions in the chemical potential, distinguished by differences in the number operator for each flavor. Additionally, using exact diagonalization (ED), we showed that the phases are independent of the sum of the chemical potentials.

In (2+1)D, no analytical predictions were available for the critical points; thus, we restricted ourselves to small systems involving 10 qubits in our proof-of-concept demonstration. These systems remain tractable via ED for lattice sizes of $2\times 2$, which we used to benchmark and verify our quantum simulations. While current limitations in circuit depth and hardware noise lead to small discrepancies in number-operator differences and larger deviations in the measured energies, our results demonstrate the feasibility of capturing essential features of level crossings, including critical points, on small lattice sizes.

In the future, we aim to investigate the scalability of our method to larger lattices using advanced adaptive ansätze such as (SC)-ADAPT-VQE~\cite{Farrell2024} and $\text{(SC)}^2$-ADAPT-VQE~\cite{Gustafson2024}, since the scaling of the proposed VQE ansatz remains prohibitive on current quantum devices. This direction is particularly relevant since the location of critical points is expected to shift with increasing system size due to finite-size effects, as observed in the (1+1)D case~\cite{Schuster2024}. Variational Monte Carlo studies of (2+1)D QED at finite density~\cite{Bender2023} indicate that $8\times 8$ lattices suffice to avoid finite-size effects at smaller couplings, whereas the required system size for the couplings considered here remains an open question. 

All data asscociated with this work is available at \cite{rosanowski_2026_20444068}.

\begin{acknowledgments}
We wish to thank Tobias Hartung, Stefan Kühn, Paolo Stornati, and Ivano Tavernelli for valuable discussions and feedback on the manuscript. The authors gratefully acknowledge the granted access to the Marvin cluster hosted by the University of Bonn.
We acknowledge the use of IBM Quantum services for this work. The views expressed are those of the authors, and
do not reflect the official policy or position of IBM or the IBM Quantum team.
This project was funded by the Deutsche Forschungsgemeinschaft (DFG, German Research Foundation) as part of the CRC 1639 NuMeriQS -- project no.\ 511713970 and under Germany's Excellence Strategy – Cluster of Excellence Matter and Light for Quantum Computing (ML4Q) EXC 2004/1 – 390534769. This work is supported with funds from the Ministry of Science, Research and Culture of the State of Brandenburg within the Center for Quantum Technology and Applications (CQTA). This work is part of the Quantum Computing for High-Energy Physics (QC4HEP) working group. This work is supported by funds from the European Union’s Horizon Europe Frame-work Programme (HORIZON) under the ERA Chair scheme with grant agreement no. 101087126.
\begin{center}
    \includegraphics[width = 0.08\textwidth]{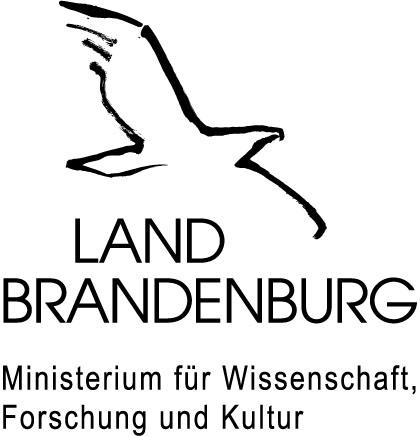}
\end{center}
\end{acknowledgments}

\FloatBarrier

\nocite{*}

\bibliography{Refs}
\newpage
\newpage
\FloatBarrier
\appendix

\section{Convergence of Truncation}
\label{app:Trunc}

\begin{figure}
    \centering
    \includegraphics[width=\linewidth]{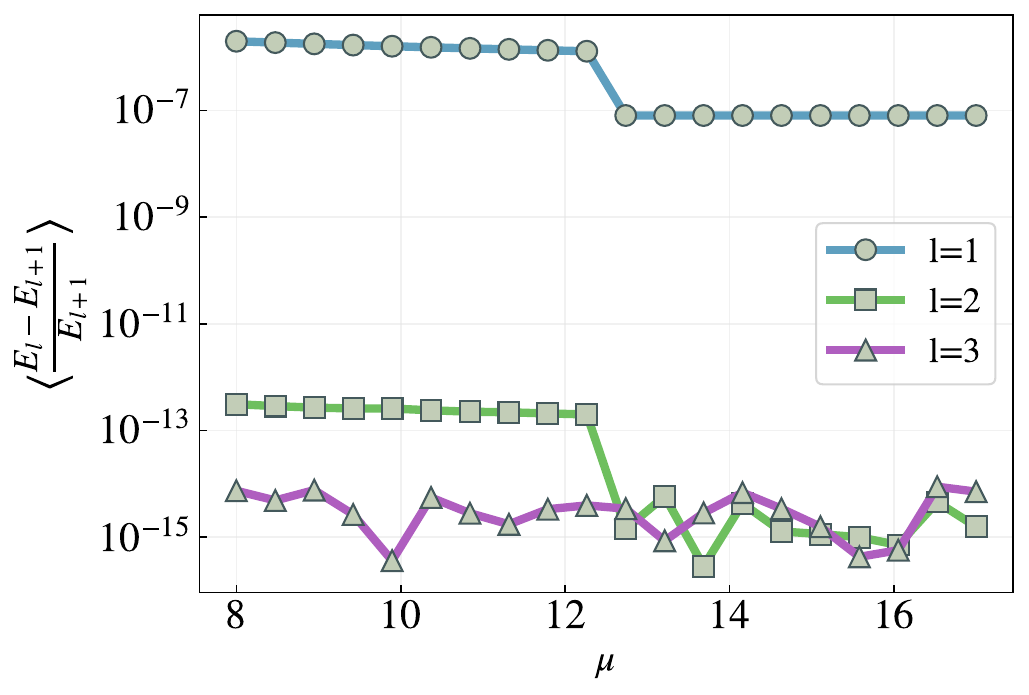}
    \caption{Relative difference $\langle (E_l-E_{l+1})/E_{l+1}\rangle$ of the ground-state energy for different levels of the truncation as a function of the chemical potential $\mu$.}
    \label{fig:trunc}
\end{figure}

In order to reduce the depth of the circuit, only the smallest truncation of $l=1$ is applied. To study the effect of this truncation on the physical results, the relative error of the ground-state energy with respect to the next higher truncation is calculated using exact diagonalization. This is shown in \cref{fig:trunc}. It can be observed that for the given system, the relative error reaches machine precision already for $l=3$. Even at $l=1$, only small deviations can be seen.

\begin{figure}
    \centering
    \includegraphics[width=\linewidth]{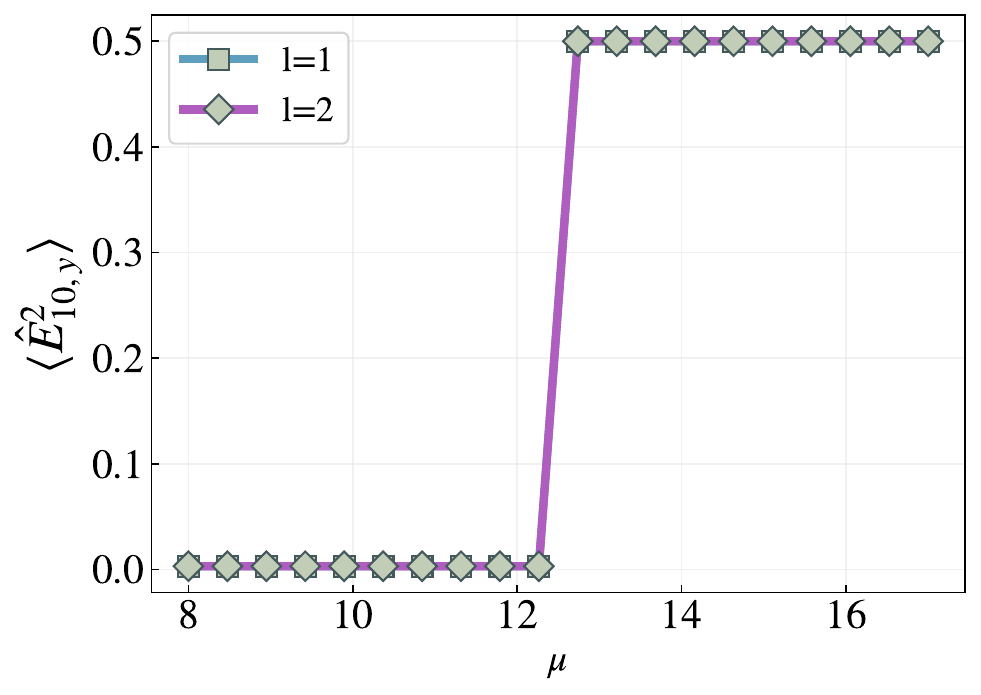}
    \caption{Expectation value of the squared electric field on the dynamical link in the system, $\langle\hat{E}_{10,y}^2\rangle$, against the chemical potential $\mu$. The data is obtained with two different truncation levels of $l=1,2$.}
    \label{fig:trunc_esq}
\end{figure}

Furthermore, the expectation value of the squared electric field on the remaining dynamical link, $\langle \hat{E}_{10,y}^2 \rangle$, as a function of the chemical potential $\mu$ (see \cref{fig:trunc_esq}), shows that even for truncations $l=1,2$ the maximum value attained is $\langle \hat{E}_{10,y}^2 \rangle = 0.5$ (up to deviations of order $10^{-5}$). A more detailed analysis for $l=1,2,3,4$ reveals that, for $\mu > \mu_{\rm LC}$, the ground state is always an equal superposition of two states with $\langle E_{10,y}^2 \rangle = 1$ and $\langle E_{10,y}^2 \rangle = 0$, respectively (up to contributions with amplitudes smaller than $10^{-7}$). This implies that, in the parameter regime considered in this work, a truncation as small as $l=1$ is sufficient to accurately describe the ground states.
Thus, this truncation provides results close to the full $U(1)$ theory for the parameter regimes used in this work.

\section{Phases with $N_f=3$ Flavors of Fermions}
\label{app:ThreeFermions}

\begin{figure}
    \centering
    \includegraphics[width=\linewidth]{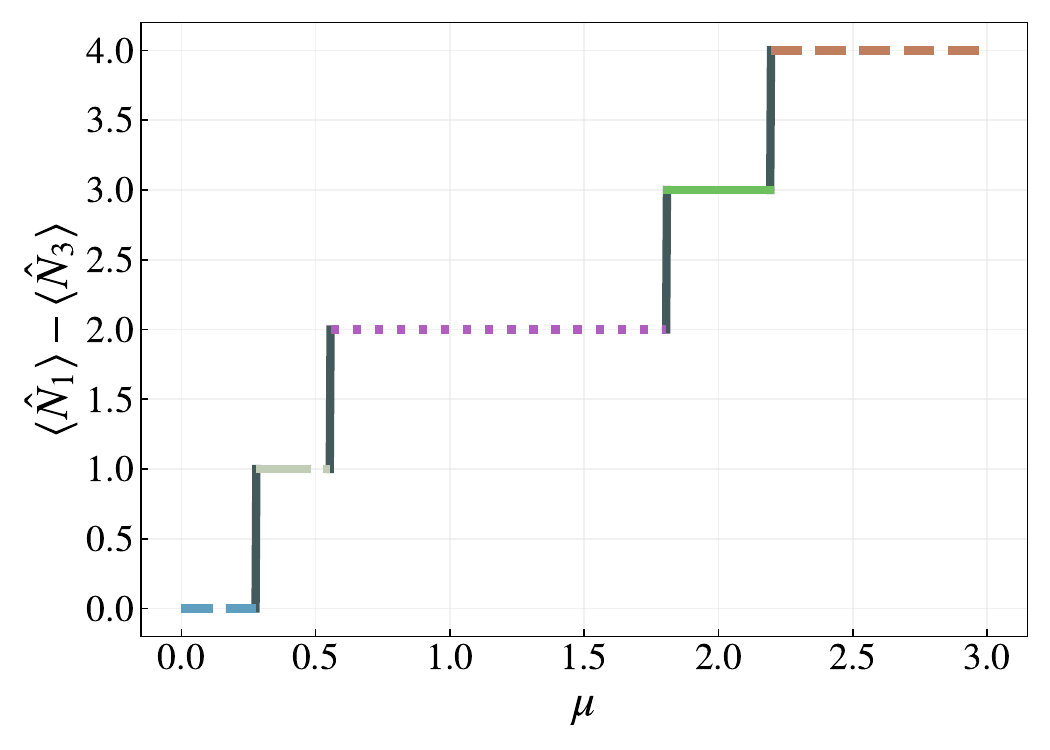}
    \caption{Difference in particle number $\langle\hat{N}_1-\hat{N}_3\rangle$ in a system with three flavors of massless fermions against the chemical potential $\mu$, where $\mu_1=-\mu$, $\mu_2=2$ and $\mu_3=\mu$ at $g=2$.}
    \label{fig:ThreeFermionsNumbers}
\end{figure}

\begin{figure}
    \centering
    \includegraphics[width=\linewidth]{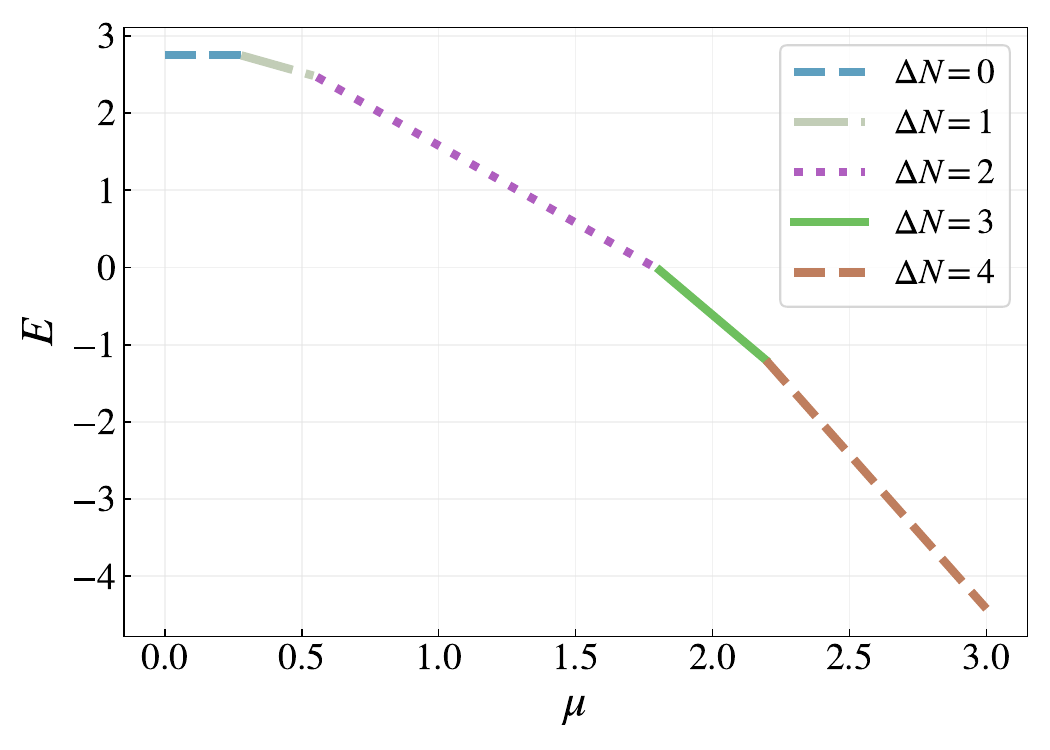}
    \caption{Energy $E$ in a system with three flavors of massless fermions against the chemical potential $\mu$, where $\mu_1=-\mu$, $\mu_2=2$ and $\mu_3=\mu$ at $g=2$. The different regions according to the difference in particle number $\Delta N=\langle\hat{N}_1-\hat{N}_3\rangle$ are colored as in \cref{fig:ThreeFermionsNumbers}.}
    \label{fig:ThreeFermionsEnergy}
\end{figure}

Although the extension to $N_f=3$ flavors of fermions is not useful for simulations on currently available quantum hardware, results using exact diagonalization can be obtained. The chemical potentials are chosen as $\mu_1=-\mu$, $\mu_2=2$, and $\mu_3=\mu$, while the coupling is set to $g=2$. The difference in particle numbers between fermions of flavors one and three now changes in steps of one instead of two as can be seen in \cref{fig:ThreeFermionsNumbers}. Consequently, there are five different phases visible now, which are also labeled by their color. These phases can also be identified as linear regions in the energy as shown in \cref{fig:ThreeFermionsEnergy}. Thus, the same qualitative behavior as in Ref.~\cite{Schuster2024} is found. Although currently available quantum hardware is not suitable for this system yet, future work might consider performing inference runs for this as well.

\section{Finite-Size Effects}
\label{app:FSE}

Throughout this work, we focus on $2\times2$ lattices with open boundary conditions, since larger systems are not yet accessible on current quantum hardware. In this setup, only a single gauge link remains dynamical after imposing Gauss’s law. Nevertheless, larger lattices can still be investigated using ED. In the following, we consider a $2\times2$ lattice with open and periodic boundary conditions as well as $2\times3$ and $2\times4$ lattices with open boundary conditions. These systems contain one, five, two, and three dynamical links, respectively.

\begin{figure}
    \centering
    \includegraphics[width=\linewidth]{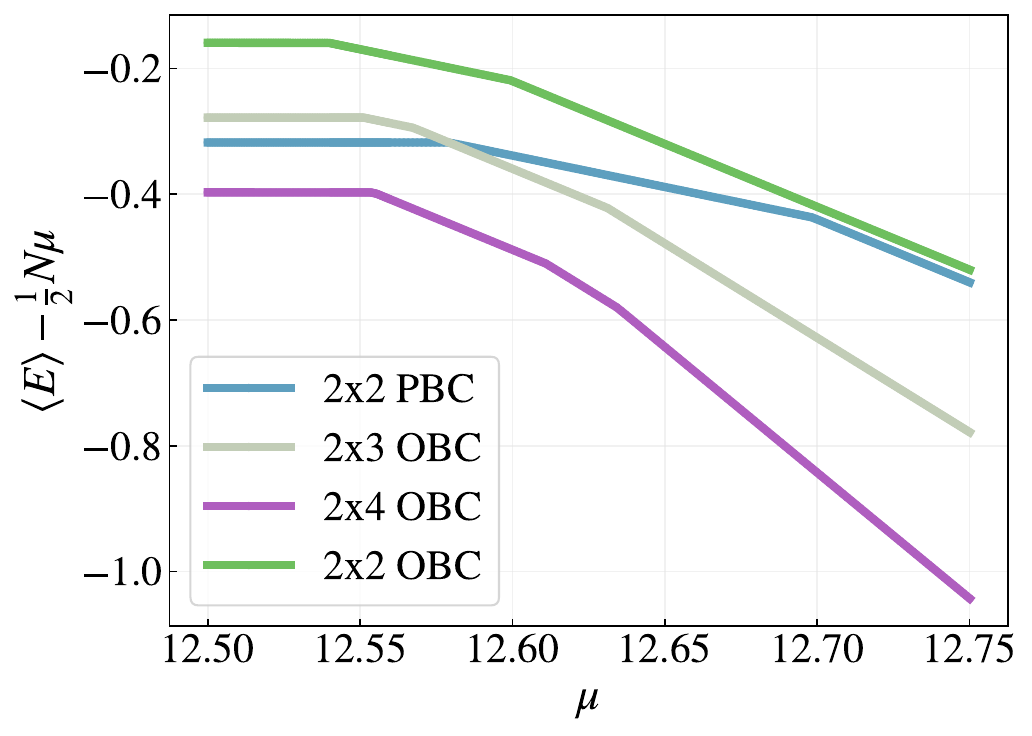}
    \caption{Expectation value of the Hamiltonian $\langle E\rangle$ as a function of the chemical potential $\mu$ for systems with $N_f=2$ fermion flavors. For better visibility, all curves are shifted by subtracting $\frac{1}{2}N\mu$, where $N=N_1+N_2=\mathrm{const.}$ within each phase. All results are obtained for truncation $l=1$, with lattice sizes and boundary conditions as indicated in the legend.}
    \label{fig:largerLatticeEnergy}
\end{figure}

The corresponding ground-state energy $\langle E\rangle$ as a function of the chemical potential $\mu$ is shown in \cref{fig:largerLatticeEnergy}. The same qualitative behavior as in \cref{fig:ClassSimEnergy} is observed. At the same time, the larger systems exhibit additional kinks, reflecting the fact that more fermions can be accommodated and therefore more level crossings appear.

\begin{figure}
    \centering
    \includegraphics[width=\linewidth]{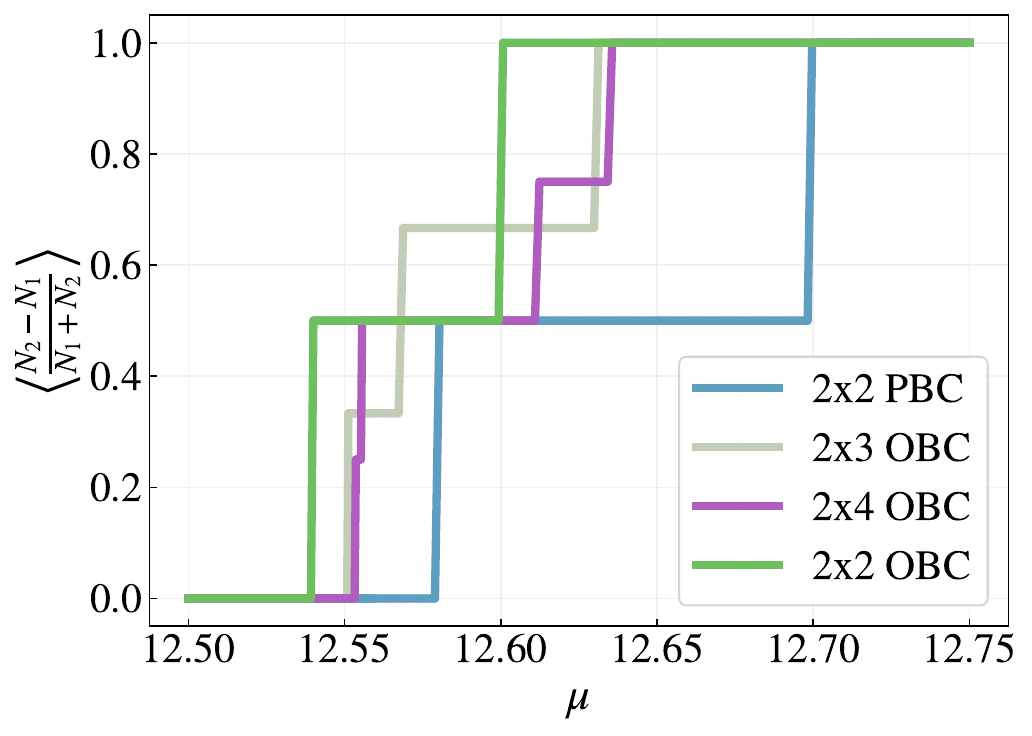}
    \caption{Ratio of the particle-number difference to the total particle number, $\big\langle \frac{N_1-N_2}{N_1+N_2}\big\rangle$, as a function of the chemical potential $\mu$ for various system sizes at truncation $l=1$. The lattice sizes and boundary conditions are indicated in the legend.}
    \label{fig:largerLatticeNumber}
\end{figure}

Similarly, \cref{fig:largerLatticeNumber} shows the ratio of the particle-number difference to the total particle number, $\big\langle \frac{N_1-N_2}{N_1+N_2}\big\rangle$, as a function of $\mu$ for the same systems. As in \cref{fig:ClassSimNum}, a step-like structure is observed. For the larger lattices, however, additional plateaus appear due to the larger number of accessible fermionic configurations. 

Overall, the larger lattices show the same qualitative behavior as the $2\times2$ system with open boundary conditions. While the positions and number of level crossings depend on the system size and boundary conditions, the characteristic step-like behavior in the particle-number observables persists. This indicates that the small-volume system studied in the main text already captures the qualitative structure of the finite-density level crossings also present on larger lattices.

\section{Leakage out of the Gauge-Invariant Sector}
\label{app:Leakage}

The Gray-encoding circuit shown in \cref{fig:GaugeCircuit} restricts the accessible states to the physical gauge-invariant subspace on an ideal quantum device. On real hardware, however, noise can induce transitions into states that are excluded by construction in the noiseless case. Although hardware errors affect the results in several ways, this specific type of leakage can be quantified directly.

\begin{figure}
    \centering
    \includegraphics[width=0.5\textwidth]{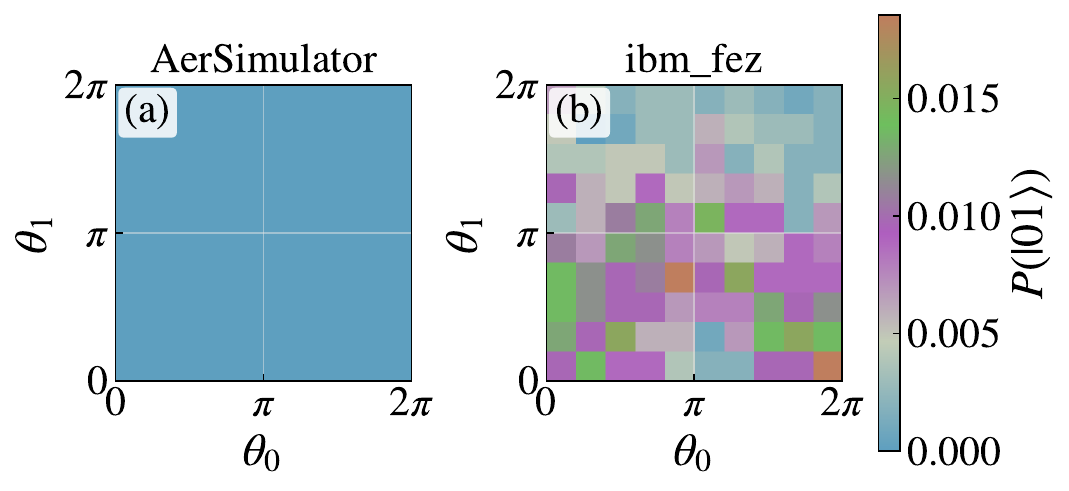}
    \caption{Probability of measuring the excluded gauge-link state $|01\rangle$ in the Gray encoding as a function of the variational parameters $\theta_0$ and $\theta_1$. The left panel shows the result obtained in a noiseless simulation with \texttt{AerSimulator}, while the right panel shows the corresponding data measured on \texttt{ibm\_fez}. In both cases, $n_\mathrm{shots}=1,\!024$ shots are used.}
    \label{fig:Leakage}
\end{figure}

To this end, \cref{fig:Leakage} shows the probability of measuring the excluded state $P\!\left(|01\rangle\right)$ as a function of the two variational parameters $\theta_0$ and $\theta_1$. We compare noiseless simulations performed with \texttt{AerSimulator} to data obtained on the \texttt{ibm\_fez} device. In both cases, $n_\mathrm{shots}=1,\!024$ shots are used. As expected, the excluded state is not populated in the noiseless simulation. On quantum hardware, by contrast, a non-zero leakage probability is observed, reaching values of up to approximately $2\%$. This behavior is consistent with the noise levels of current NISQ devices.

\section{Zero-Noise Extrapolation}
\label{app:ZNEexplicit}

\begin{figure*}
    \centering
    \includegraphics[width=\textwidth]{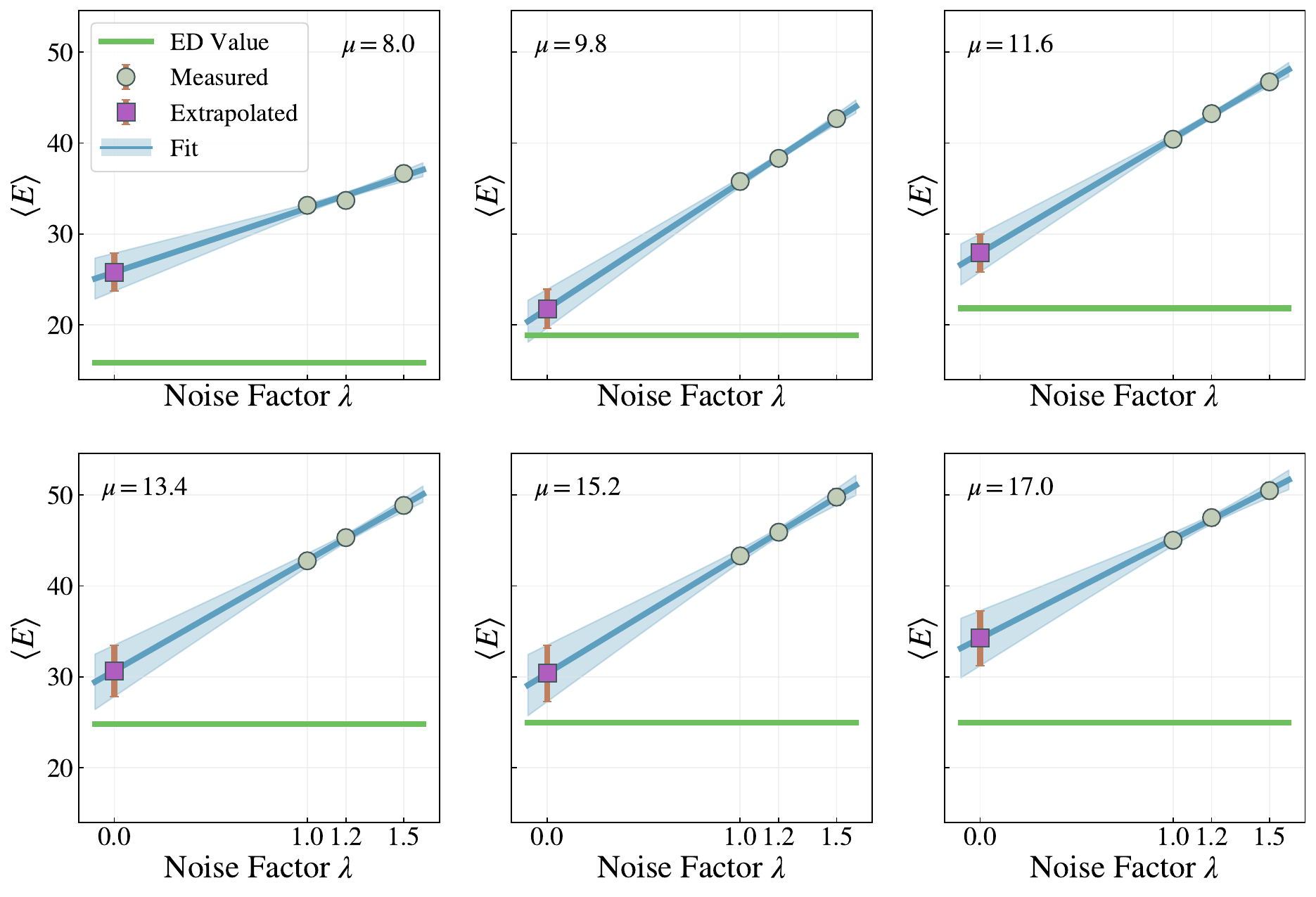}
    \caption{ZNE for the observable $\langle E\rangle$ with the three noise factors $\lambda\in[1,1.2,1.5]$. Extrapolation is performed with a linear fit to the data points measured on \texttt{ibm\_marrakesh} with $8,\!192$ shots. The different panels correspond to the different values of $\mu$ being used.}
    \label{fig:ZNEexplicit}
\end{figure*}

The ZNE has been used in \cref{subsec:InferenceRun} to reduce the influence of hardware errors.
We applied partial gate folding to achieve noise factors of $\lambda\in[1,1.2,1.5]$. The extrapolation is done using a linear fit as this described the data points best.
Here, in \cref{fig:ZNEexplicit} the extrapolation for the energy is explicitly shown. The different panels correspond to the different values of $\mu\in[8,9.8,11.6,13.4,15.2,17]$ being used in \Cref{fig:InferenceEnergy}.  Since $n_\text{shots}=8,\!192$ shots have been used, the uncertainties on the data points are smaller than the markers.

\end{document}